# Ultra low loss broadband 1 × 2 optical power splitters with various splitting ratios

KIYANOUSH GOUDARZI,[1] SUNWOOK KIM,[1] IKIMO PARK,[2] AND HAEWOOK HAN[1,*]

[1]Department of Electrical Engineering, Pohang University of Science and Technology, Pohang 37673, Republic of Korea
[2]Department of Electrical and Computer Engineering, Ajou University, 206 Worldcup-ro, Youngtong-gu, Suwon 16499, Republic of Korea.
*hhan@postech.ac.kr

**Abstract:** We designed Si-based all-dielectric 1 × 2 TE and TM power splitters with various splitting ratios and simulated them using the inverse design of adjoint and numerical 3D finite-difference time-domain methods. The proposed devices exhibit ultra-high transmission efficiency above 98 and 99%, and excess losses below 0.1 and 0.035 dB, for TE and TM splitters, respectively. The merits of these devices include a minor footprint of 2.2 × 2.2 µm$^2$ and a broad operating bandwidth of 200 nm with a center wavelength of $\lambda$ = 1.55 µm.

## 1. Introduction

All-dielectric metamaterials (MMs) contain artificial structures that steer and confine light. Dielectric materials with high refractive indices are good candidates for designing MM devices. Among the best candidates for this purpose is silicon, with its prominent characteristics including high refractive index (nearly 3.48) in a 1.55 µm communication window, high refractive index contrast between Si and SiO$_2$, and compatibility with CMOS fabrication technology. Another feature of Si is its capability to steer and confine incident light in the $\lambda/n$ subwavelength where $\lambda$ and $n$ denote the wavelength of incident light and refractive index, respectively; these jointly contribute to the application of silicon in high-density optical integrated circuits (OICs). Owing to the benefits of Si-based MMs, numerous optical devices based on MMs have been designed and fabricated to date [1–3].

For designing an optical component, numerous structural simulations are required to find an optimized structure, which enforce high computational costs and time-consuming simulations. To overcome these problems, and design efficient and compact structures, researchers have developed inverse design methods, including optimizations and artificial intelligence (machine/deep learning). Various optical structures have been designed by artificial intelligence, namely all-dielectric MMs [4], optical scattering unit [5], nanoparticles [6, 7], multilayers [8, 9], metasurfaces [10–13], and chiral MMs [14]. Currently, inverse design based on optimization has attracted attention as an efficient method. Furthermore, a multitude of optical devices have been simulated and fabricated according to these methods, among which are optical power splitters (OPSs) [4, 15–20], wavelength demultiplexers [21–23], polarization splitters [24, 25], and couplers [21].

There are two major optimization methods for optical design, namely adjoint [21–23, 26] and direct binary search [17, 18, 24]. The major differences in between are simulation time and efficiency of the optimized devices. The optimization process time and efficiency of the adjoint method are lower and higher, respectively, compared to the direct binary search method. Direct binary search optimizes structures, which can be modeled as zero and one states. In contrast, the adjoint method is capable of being applied to design myriads of structures, including binary [16] and other structures [21–23]. In this study, TE and TM OPSs with various splitting ratios were designed and simulated employing the adjoint method. The proposed devices exhibit great application potential owing to their small footprint, high transmission efficiency, broad bandwidth, and low computational time requirement.

## 2. Theory and design

Designing an OIC requires a multitude of basic components such as OPSs, polarization demultiplexers, filters, and wavelength demultiplexers. Although OPSs are basic elements in OICs, those with arbitrary splitting ratios are applicable in signal processing, optical equalization, and feedback circuits [27]. Here, OPSs have been proposed and designed using an efficient and powerful adjoint method, which incorporates a gradient descent algorithm for the optimization. In this method, first proposed by Miller to design solar cell structures [28], the structure shape is altered to obtain the desired outputs. It contains three stages, namely, grayscale, binarization, and design for manufacturing [21]. In the grayscale phase, the structure's permittivity is varied between two predefined permittivity values, $\varepsilon_{max}$ and $\varepsilon_{min}$. The binarization stage uses the Heaviside function to change these values either $\varepsilon_{max}$ or $\varepsilon_{min}$. In the design for the manufacturing process, fabrication constraints such as minimum shape curvature are applied. This method also requires an initial structure region for optimization; to this end, a simple structure has been defined in this study for the OPSs according to Fig. 1. The desired output or figure of merit (FOM) of this method is defined as [16, 26]:

$$\text{FOM} = \frac{1}{4} \frac{\left| \int_{S'} [\mathbf{E}(\mathbf{r}') \times \mathbf{H}^*(\mathbf{r}') + \mathbf{E}^*(\mathbf{r}') \times \mathbf{H}(\mathbf{r}')] \cdot d\mathbf{S}' \right|^2}{\int_{S'} \text{Re}[\mathbf{E}_0^*(\mathbf{r}') \times \mathbf{H}_0^*(\mathbf{r}')] \cdot d\mathbf{S}'} \quad (1)$$

where $\mathbf{E}$ and $\mathbf{H}$ denote the electric and magnetic field vectors at $S'$ (cross-section of the output branches), and $\mathbf{E}_0$ and $\mathbf{H}_0$ are electric and magnetic field vectors of the incident wave, respectively. Furthermore, a small variation of permittivity in the initial structure, which is the space with the dimensions of $L \times L \times T$ as shown in Fig. 1, will induce the electric dipole moment, leading to electromagnetic field variation at $\mathbf{r}'$ (the position at the $S'$) surface.

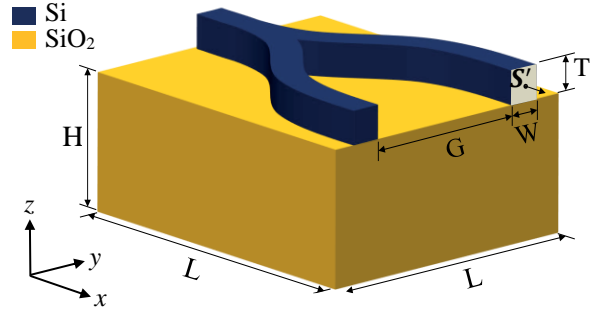

Fig.1. Schematic view of the initial device structure.

## 3. Simulation results and discussions

In this section, both 1 × 2 OPSs with $TE_{00}$ and $TM_{00}$ modes, respectively, with their electrical fields in $y$- and $z$- directions and various splitting ratios are discussed. The prototype structure for both TE and TM OPSs is shown in Fig. 1. The figure shows 1 × 2 OPSs with one input and two output Si waveguide ports with a refractive index of 3.48 and 0.4 µm waveguide width. The footprint of the optimization region is 2.2 × 2.2 µm² with a thickness of 0.4 µm. The material includes two branches of Si, and the wavelength for the devices is in the 1450–1650 nm range (200 nm bandwidth). The structural parameters of $L$, $W$, $G$, $H$, and $T$ have been chosen as 2.2, 0.4, 1.2, 3, and 0.4 µm, respectively.

For optimization processes and simulations of the structure, the 3D FDTD module of Lumerical software was employed. Furthermore, the Python embedded in FDTD module of Lumerical software has been applied in the optimization process. The mesh sizes in $x$-, $y$-, and $z$- directions for 3D FDTD simulations have been set to $dx = 20$ nm, $dy = 20$ nm, and $dz = 50$ nm, respectively.

The simulation results of the optimized TE and TM OPSs with splitting ratios of 50:50, 60:40, 70:30, and 80:20 are shown in Figs. 2 and 3, correspondingly. In the following figures, UB and LB stand for upper and lower branches, respectively. The simulation time for optimizing the OPSs is roughly 2 h per device with the computational resource being a PC with 3 GHz Core-i9 CPU and 128 GB RAM.

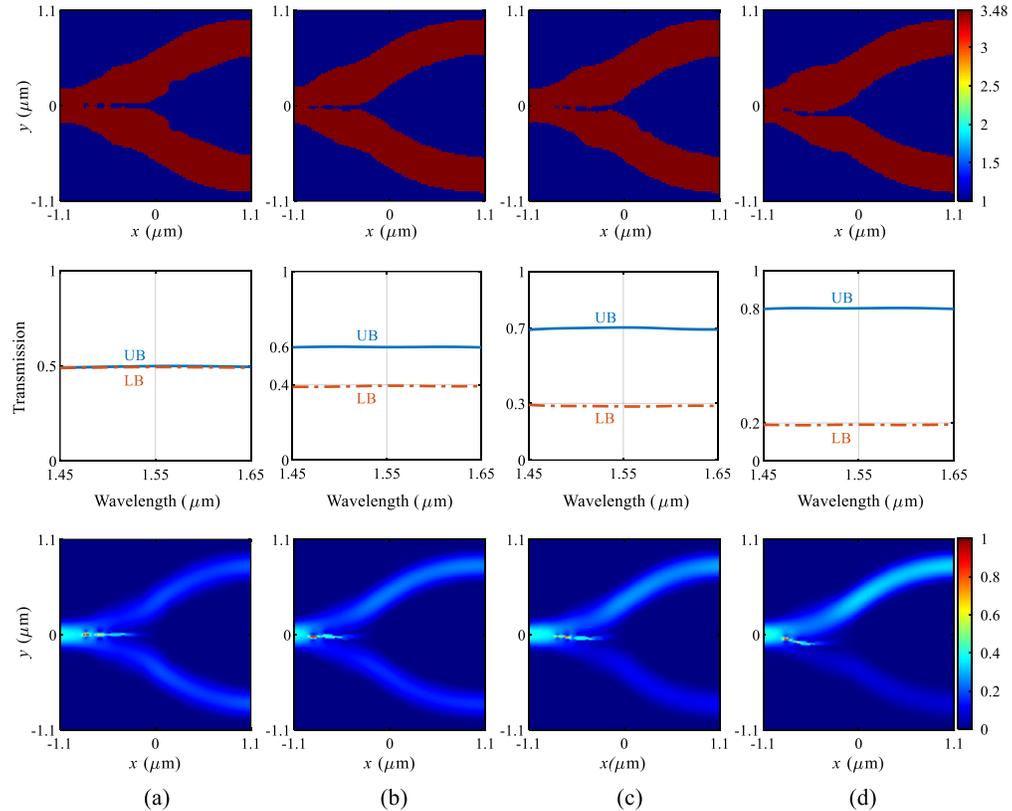

Fig. 2. Simulation results for the optimized 1 × 2 TE OPSs. Columns (a)–(d) show the devices with splitting ratios of 50:50, 60:40, 70:30, and 80:20, respectively. The rows (up to down) depict refractive index profile, transmission spectrum, and the normalized power distribution at $\lambda = 1.55$ µm, respectively.

The total transmission efficiency at both outputs is defined as $(T_1 + T_2)$, where $T_1$ and $T_2$ denote the transmissions at each output port. As illustrated in Fig. 2, the efficiencies are 98.8, 99.6, 98.23 and 99.11% for TE OPSs with splitting ratios of 50:50, 60:40, 70:30, and 80:20, respectively. Fig. 2 indicates that based on the boundary conditions of Maxwell equations, the narrow gap between the output branches in TE mode structures results in discontinuity of the electrical field and subsequently causes electrical field amplification in the gaps. According to Fig. 3, the overall transmission efficiency for TM OPSs with splitting ratios of 50:50, 60:40, 70:30, and 80:20 are 99, 99.59, 99.65, and 99.38%, respectively.

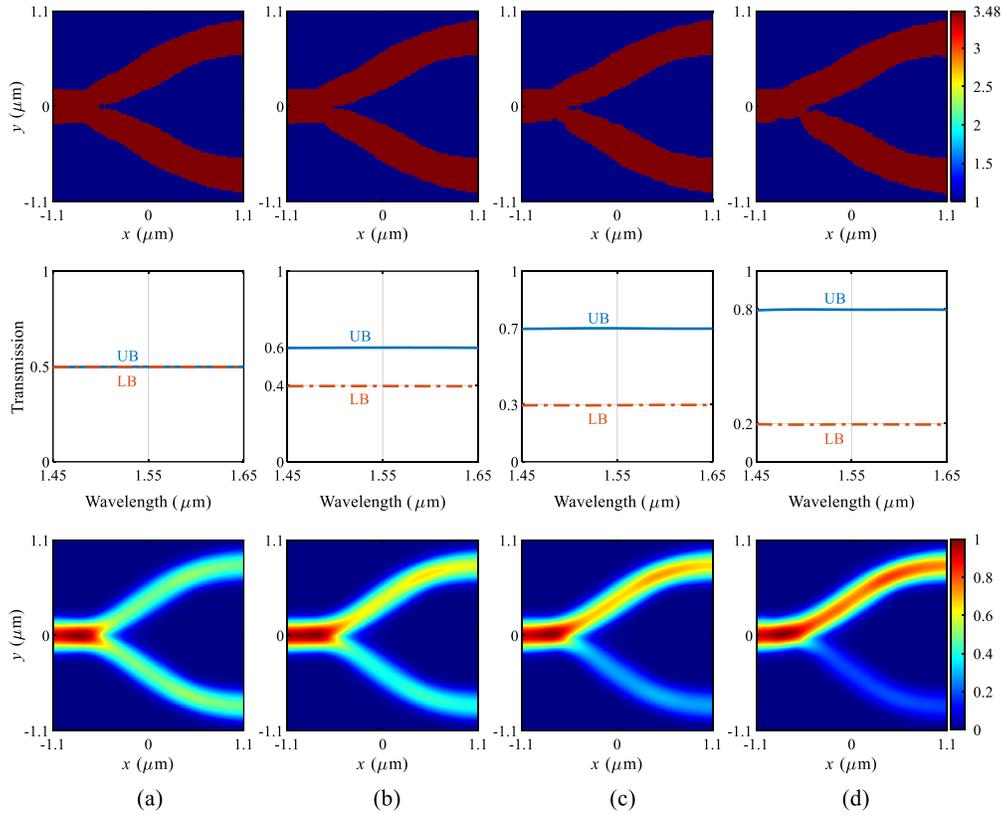

Fig. 3. Simulation results for the optimized 1 × 2 TM OPSs. Columns (a)–(d) show the devices with splitting ratios of 50:50, 60:40, 70:30, and 80:20, respectively. The rows (up to down) depict refractive index profile, transmission spectrum, and the normalized power distribution at $\lambda$ = 1.55 μm, respectively.

Figure 4 shows the excess loss parameter for the proposed TE and TM OPSs; it is evident that the excess loss is below 0.1 dB for TE and 0.035 dB for TM OPSs. Based on the simulations, the optimized designed structures have ultra-high transmission efficiency, a simple and small structure, fabrication feasibility and very short optimization time. In general, three factors aid in the determination of the optimization time, namely, the structure size, proper selection of initial structure for optimization, and selection of an efficient algorithm. The physical phenomenon behind these structures is that the optimized region, which is a Y splitter, acts as a coupler including one input and two output waveguides. The incident light with the specific guided mode is coupled to the output waveguides; the portion of coupled optical waves depends on the geometry of output branches; i.e., based on Snell's law, the incident light from the input waveguide is transmitted and reflected to the output waveguides, which in turn depends on the topology of output branches, which realizes different splitting ratios. Notably, the guided waves are transmitted through the devices based on total internal reflections. In comparison to the other studies [4, 15–21], the proposed OPSs exhibit the highest transmission efficiency (> 98 %) and the shortest simulation time. Further, the proposed OPSs have the smallest footprint, broadband, and exhibit the widest bandwidth (almost 200 nm) compared to the other studies except Ref. [21]. Despite the wider bandwidth obtained in Ref. [21], the optimized OPSs have a very flat bandwidth spectrum. Moreover, these devices have been designed with different splitting ratios of 50:50, 60:40, 70:30 and 80:20 in either $TE_{00}$ or $TM_{00}$ modes. The other major differences between the optimized OPSs and other studies is the structure, as the proposed OPSs have only two branches of Si waveguides to facilitate the fabrication process.

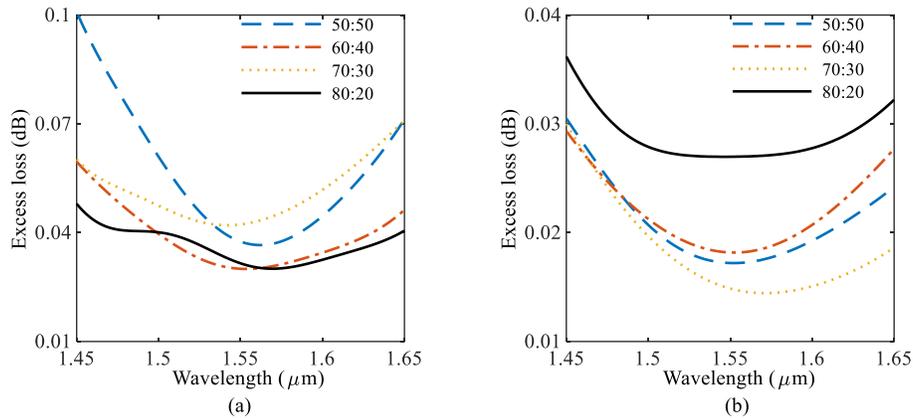

Fig. 4. Excess loss for (a) TE, and (b) TM OPSs with respect to wavelength at splitting ratios of 50:50, 60:40, 70:30, and 80:20.

## 4. Conclusion

We designed and simulated $1 \times 2$ TE and TM OPSs with different splitting ratios using the adjoint method. The optimized TE and TM OPSs with splitting ratios of 50:50, 60:40, 70:30, and 80:20 yield high efficiency (> 98% for $TE_{00}$ and > 99% for $TM_{00}$ modes), low excess loss (below 0.035 and 0.1 dB for $TM_{00}$ and $TE_{00}$ modes, respectively), a miniscule footprint ($2.2 \times 2.2$ µm$^2$), very short optimization time (2 h per each OPS), broad and flat bandwidth (200 nm), operation in either $TE_{00}$ or $TM_{00}$ modes, and simple optimized structures. These silicon-on-insulator (SOI) devices are compatible with CMOS fabrication technology. The adjoint optimization method and 3D FDTD module of Lumerical software have been adopted for designing and simulating the OPSs. Overall, the abovementioned features present the proposed devices as excellent candidates for OICs. Therefore, these optimized devices will be applicable in the mass production of high-density, high-efficiency, and high-speed OICs in the future.

**Acknowledgments.** This work was supported by MSIT (Ministry of Science and ICT), Korea, under the ICT Creative Consilience program (IITP-2020-2011-1-00783) supervised by the IITP (Institute for Information & communications Technology Planning & Evaluation).